\documentclass[11pt,a4paper]{article}

\usepackage[T1]{fontenc}
\usepackage[utf8]{inputenc}
\usepackage[english]{babel}
\usepackage{lmodern}
\usepackage[a4paper,margin=24mm,headheight=14pt,headsep=7mm,footskip=11mm]{geometry}
\usepackage{microtype}
\usepackage{amsmath,amssymb}
\usepackage{graphicx}
\usepackage{xcolor}
\usepackage{array,booktabs,multirow}
\usepackage[font=small,labelfont=bf,labelsep=period]{caption}
\usepackage{subcaption}
\usepackage{placeins}
\usepackage{overpic}
\usepackage{titling}
\usepackage{authblk}
\usepackage{titlesec}
\usepackage{fancyhdr}
\usepackage[numbers,sort&compress]{natbib}
\usepackage[unicode=true,colorlinks=true,allcolors=black]{hyperref}

\hypersetup{
  pdftitle={In situ performance of the Vertex Silicon Plane in the BM@N experiment during Run 9},
  pdfauthor={D. Dementev, M. Shitenkov, A. Kolozhvari, V. Leontyev, A. Sheremetev, D. Andreev, R. Arteche Diaz, D. Baranov, C. Ceballos Sanchez, I. Gabdrakhmanov, I. Gorelikov, T. Lygdenova, I. Osokin, S. Reyes Pe\~na, I. Rufanov, E. Tsapulina, A. Voronin, Yu. Murin},
  pdfsubject={In situ performance of the BM@N Vertex Silicon Plane during Run 9},
  pdfkeywords={silicon microstrip detectors, particle tracking detectors, solid-state detectors, detector performance, BM@N, NICA},
  pdflang={en-US}
}
\titleformat{\section}{\large\bfseries}{\thesection}{0.7em}{}
\titleformat{\subsection}{\normalsize\bfseries}{\thesubsection}{0.7em}{}
\titleformat{\subsubsection}{\normalsize\itshape}{\thesubsubsection}{0.7em}{}
\titlespacing*{\section}{0pt}{2.2ex plus 0.5ex minus 0.2ex}{1.0ex}
\titlespacing*{\subsection}{0pt}{1.8ex plus 0.4ex minus 0.2ex}{0.7ex}
\titlespacing*{\subsubsection}{0pt}{1.5ex plus 0.3ex minus 0.2ex}{0.5ex}

\fancypagestyle{plain}{%
  \fancyhf{}
  \fancyfoot[C]{\small\thepage}
  
}

\pretitle{\begin{center}\LARGE\bfseries}
\posttitle{\par\end{center}\vspace{0.2em}}
\date{}
\predate{}
\postdate{}

\renewenvironment{abstract}
  {\par\small\noindent\textbf{Abstract}\par\vspace{0.4em}\noindent\ignorespaces}
  {\par\normalsize\vspace{0.6em}}
\newenvironment{keyword}
  {\par\small\noindent\textbf{Keywords: }\ignorespaces}
  {\par\normalsize\vspace{0.8em}}
\newcommand{\panelmark}[1]{\par\vspace{2pt}{\small\bfseries(#1)}}

\title{In situ performance of the Vertex Silicon Plane in the BM@N experiment during Run~9}

\author[1]{D.~Dementev\thanks{Corresponding author: \href{mailto:dementiev@jinr.ru}{\nolinkurl{dementiev@jinr.ru}}.}}
\author[1]{M.~Shitenkov}
\author[1]{A.~Kolozhvari}
\author[1,2]{V.~Leontyev}
\author[1]{A.~Sheremetev}
\author[1]{D.~Andreev}

\author[1,3]{R.~Arteche Diaz}
\author[4]{D.~Baranov}
\author[1]{C.~Ceballos Sanchez}
\author[1]{I.~Gabdrakhmanov}
\author[1]{I.~Gorelikov}
\author[1]{T.~Lygdenova}
\author[1]{I.~Osokin}
\author[1,5]{S.~Reyes Pe\~na}
\author[1]{I.~Rufanov}
\author[1]{E.~Tsapulina}
\author[1]{A.~Voronin}
\author[1]{Yu.~Murin}

\affil[1]{Veksler and Baldin Laboratory of High Energy Physics, Joint Institute for Nuclear Research, Dubna, Moscow Region, Russia}

\affil[2]{Moscow State University, Moscow, Russia}

\affil[3]{Center of Technological Applications and Nuclear Development, Havana, Cuba}

\affil[4]{Meshcheryakov Laboratory of Information Technologies, Joint Institute for Nuclear Research, Dubna, Moscow Region, Russia}

\affil[5]{Higher Institute of Technologies and Applied Sciences, Havana University, Havana, Cuba.}

\begin{document}
\maketitle

\begin{abstract}
The Vertex Silicon Plane (VSP) was commissioned and operated for the first time
as the most upstream tracking station of the Baryonic Matter at Nuclotron
(BM@N) experiment at the NICA accelerator complex during Run~9. The station
consisting of six double-sided silicon microstrip modules based on the CBM--BM@N
detector technology was installed about \(8~\mathrm{cm}\) downstream from the
target inside the SP-41 analyzing magnet.

The in situ module performance was characterized in terms of the
signal-to-noise ratio, time resolution, and Lorentz drift in the magnetic
field. The most probable cluster signal-to-noise ratio ranges from 15.5 to
33.2, while the time resolution relative to the trigger pulse is about
\(12~\mathrm{ns}\). The tracking performance was evaluated using track--hit
residuals and hit-matching efficiency; two independent reference
configurations based on the Forward Silicon Detector (FSD) and Gas Electron
Multiplier (GEM) tracking systems give consistent efficiencies of about
\(95\%\).

The contribution of the VSP to the BM@N tracking performance was further
studied through the transverse track impact parameter and the reconstruction
of short VSP--GEM tracklets. Owing to its position close to the target and its
coverage at large polar angles, the VSP provides an additional near-target
tracking point and makes it possible to extend the geometrical acceptance
beyond that of the FSD- and GEM-based central tracking system.
\end{abstract}

\begin{keyword}
Silicon microstrip detectors, particle tracking detectors, solid-state detectors, detector performance, BM@N, NICA
\end{keyword}

\section{Introduction}
\label{sec:introduction}
\subsection{The BM@N experiment}
\label{subsec:bmn_experiment}

The Baryonic Matter at Nuclotron (BM@N) is a fixed-target
experiment at the Nuclotron-based Ion Collider fAcility (NICA) in Dubna,
Russia. It is designed to study strongly interacting matter produced in
heavy-ion collisions at beam kinetic energies up to
$4.5~A\mathrm{GeV}$~\cite{bmn_physics}. The detector is optimized for
charged-particle reconstruction in the forward-rapidity region. A detailed
description of the BM@N spectrometer is given in Ref.~\cite{bmn_setup}.

The central tracking system (CTS) is located inside the SP-41 large-aperture warm dipole analyzing magnet. It
combines four Forward Silicon Detector (FSD) stations based on double-sided
silicon microstrip sensors followed by seven Gas Electron Multiplier (GEM) tracking
planes. In Run~9, the system was complemented by the Vertex Silicon Plane
(VSP), installed upstream of the FSD and GEM stations.

The VSP consists of six CBM--BM@N-type double-sided silicon strip detector
(DSSD) modules developed for the large-acceptance Silicon Tracking System of
the BM@N experiment \cite{bmn_physics}. A primary goal of the first VSP
commissioning in BM@N was to demonstrate the integration of its data-driven
readout into the BM@N data-acquisition system (DAQ). The integration was
evaluated through the time correlation of VSP hits with accepted BM@N triggers
and their spatial consistency with independently reconstructed tracks.

The VSP provides a space-point measurement close to the target
for particles emitted at polar
angles of approximately \(25^\circ\)--\(52^\circ\). This angular region is only
weakly covered by the FSD stations, whereas the VSP acceptance overlaps that
of the first GEM stations, enabling track reconstruction towards the midrapidity region (see Sec.~\ref{subsec:data_analysis}).

\begin{figure}[htbp]
\centering
\includegraphics[width=0.95\textwidth]{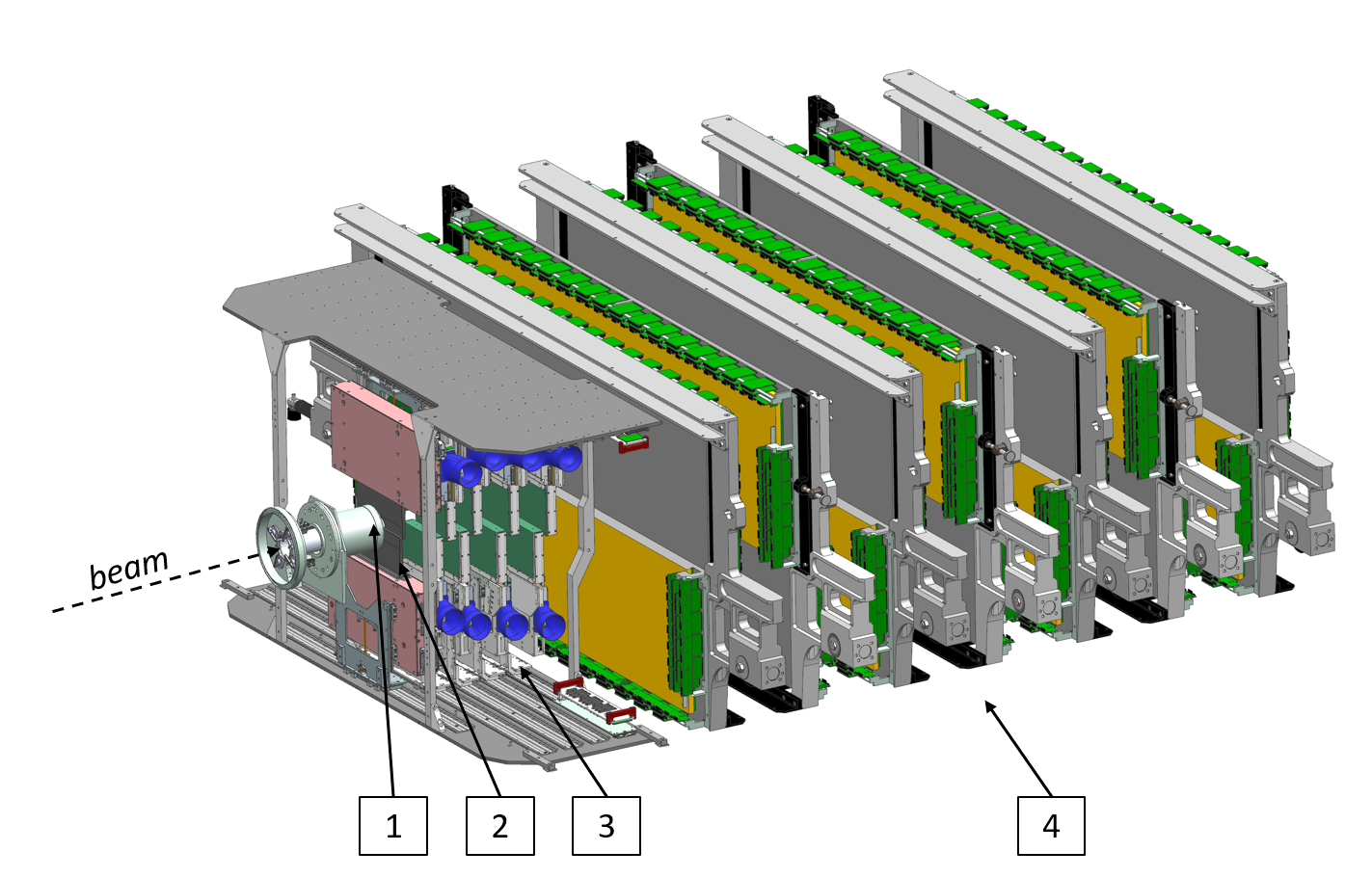}
\caption{CAD view of the configuration of the central tracking system (CTS)
for Run~9 inside the SP-41 dipole magnet. The beam direction is indicated by
the dashed arrow.
The numbered elements denote: 1 -- target chamber, 2 -- VSP station,
3 -- FSD stations, and 4 -- GEM tracking planes. The beam pipe is not shown.}
\label{fig:central_tracking_cad}
\end{figure}

Figure~\ref{fig:central_tracking_cad} shows the configuration of the central
tracking system for Run~9. The incoming beam was measured by the Silicon Beam Tracker (SiBT), located upstream of the target ~\cite{Sedykh2025SiBT,Chemezov2026SiBT}. In this analysis, the SiBT information provides an independent beam constraint used for primary-track selection and for alignment.

\FloatBarrier
The main tracking-detector parameters are summarized in
Table~\ref{tab:tracking_detector_parameters}. The FSD and GEM systems are
described in Ref.~\cite{bmn_setup}.

\begin{table}[!htbp]
\centering
\caption{Parameters of the tracking detectors of the CTS. The effective pitch
in the second coordinate is derived from the strip pitch and the corresponding
stereo angle.}
\label{tab:tracking_detector_parameters}

\begin{tabular}{lccccc}
\toprule
Detector &
Planes &
\begin{tabular}{c}
Strip pitch (X),\\
mm
\end{tabular} &
\begin{tabular}{c}
Effective pitch (Y),\\
mm
\end{tabular} &
\begin{tabular}{c}
Active area,\\
m$^2$
\end{tabular} &
\begin{tabular}{c}
Channels,\\
$\times 10^3$
\end{tabular}
\\
\midrule

VSP &
1 &
0.058 &
0.44 &
0.021 &
12 \\

FSD &
4 &
0.095 &
2.36 &
0.307 &
54 \\

GEM &
7 &
0.80 &
3.1 &
9.58 &
87 \\

\bottomrule
\end{tabular}
\end{table}

\subsection{VSP detector}
\label{subsec:vsp_detector}

\begin{figure}[htbp]
\centering
\includegraphics[
    width=0.92\textwidth,
]{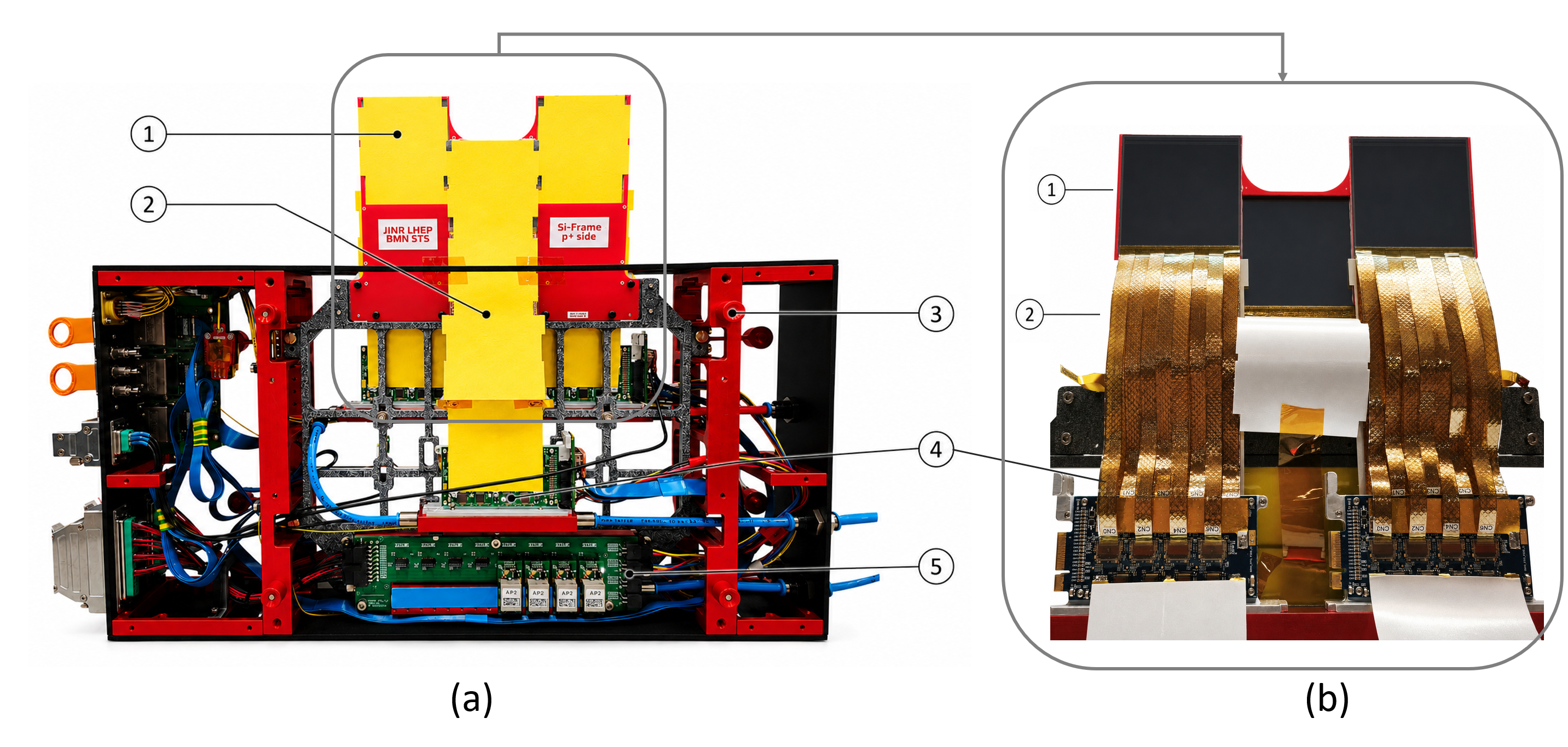}
\caption{Photographs of one VSP half-station.
(a) Half-station before installation of the final shielding covers and
positioning mechanics. (b) Enlarged view of the marked DSSD-module region with
the microcable stacks exposed. The same numbering is used in both panels:
1 -- double-sided silicon strip detector (DSSD) modules;
2 -- microcable stacks, covered by aluminum shielding;
3 -- alignment-mark seats;
4 -- front-end boards with STS/MUCH-XYTER ASICs;
5 -- power board with magnetic-field-resistant DC/DC converters.}
\label{fig:vsp_station_photo}
\end{figure}

The VSP consists of six double-sided silicon microstrip modules arranged in two
half-stations around the vacuum beam pipe. One of the half-stations is shown in
Fig.~\ref{fig:vsp_station_photo}. The sensors are located 7.4--8.5~cm
downstream from the target, covering polar angles of approximately
\(23^\circ\)--\(52^\circ\). The lower limit of the angular acceptance is
defined by the central opening for the beam pipe.

\begin{figure}[htbp]
\centering

\begin{minipage}[t]{0.48\textwidth}
    \vspace{0pt}
    \centering
    \includegraphics[width=\linewidth]{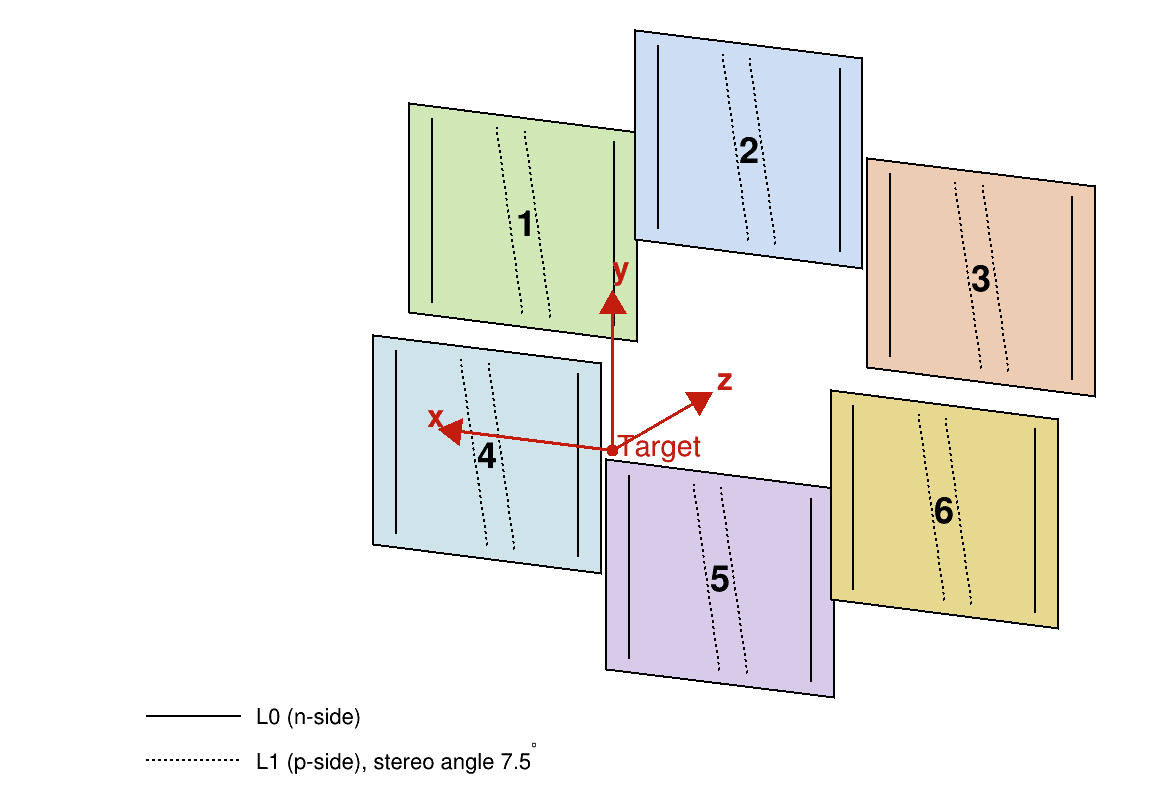}
    \panelmark{a}
\end{minipage}
\hfill
\begin{minipage}[t]{0.48\textwidth}
    \vspace{0pt}
    \centering
    \includegraphics[height=5.2cm,keepaspectratio]{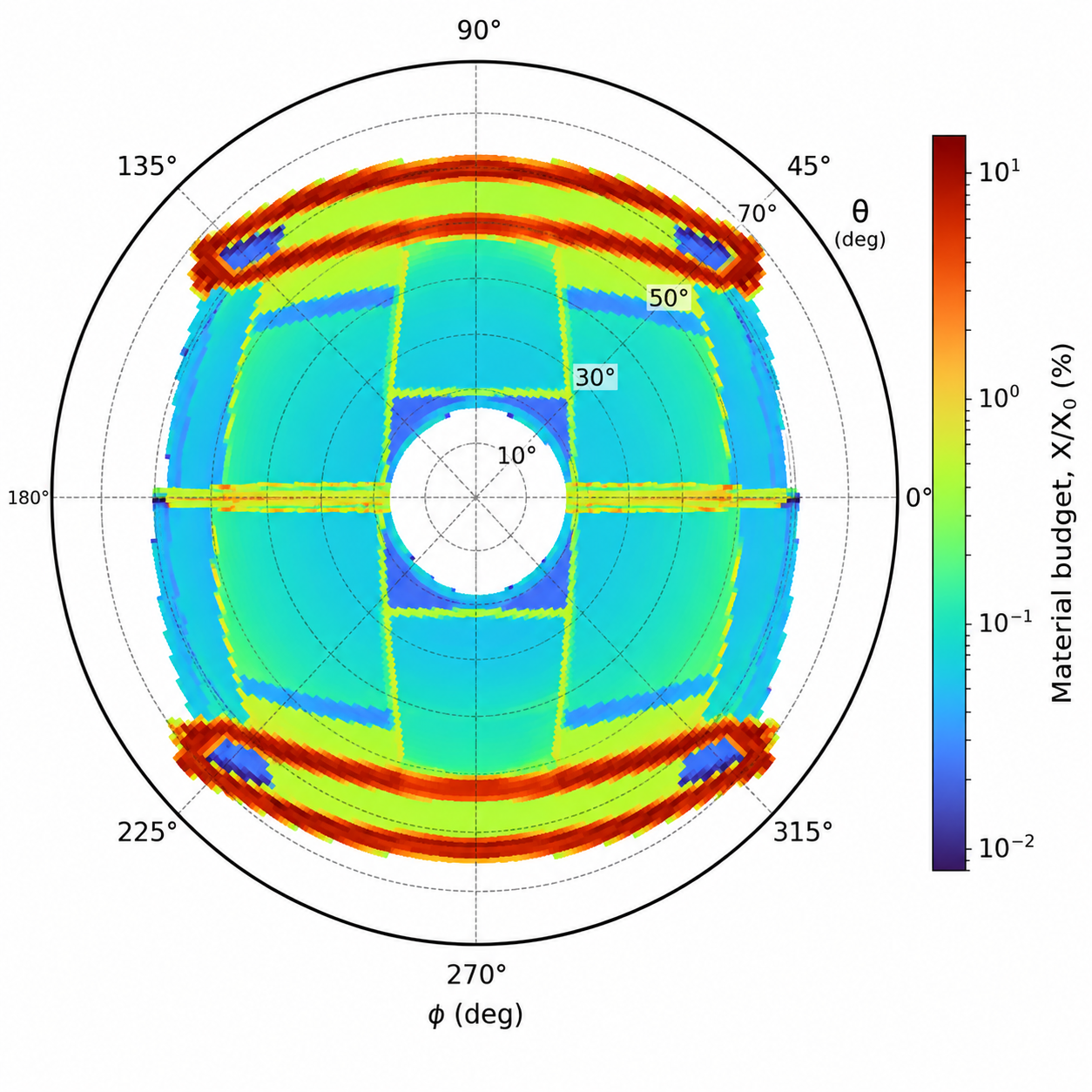}
    \panelmark{b}
\end{minipage}

\caption{VSP geometry and material distribution.
(a) Layout of the six VSP modules obtained from the Run~9 ROOT geometry. The
coordinate axes and module numbering are shown. Solid and dashed lines
schematically indicate the strip directions on the n-side (L0) and p-side
(L1), respectively.
(b) Material-budget distribution of the VSP station in polar coordinates,
including the silicon modules, module frames, and supporting structures. The
radial coordinate corresponds to the polar angle \(\theta\), while the
azimuthal coordinate is \(\phi\). The color scale shows the integrated material
budget \(X/X_0\).}
\label{fig:vsp_halfstation}
\end{figure}

The module layout and strip orientations are shown in
Fig.~\ref{fig:vsp_halfstation}(a). The average material budget in the sensitive
area is about \(0.8\%\) of a radiation length ($X_0$), with local variations due to the module frames
and support structures, as shown in Fig.~\ref{fig:vsp_halfstation}(b).

The VSP modules were assembled at JINR as part of the first batch produced for
the wide-aperture Silicon Tracking System of the BM@N experiment
~\cite{module_assembly}. Each module incorporates a double-sided silicon strip
sensor with dimensions of \(62\times62~\mathrm{mm}^2\),
a sensitive area of about \(60\times60~\mathrm{mm}^2\), and a thickness of
\(320~\mu\mathrm{m}\). Each sensor side contains 1024 strips with a pitch of
\(58~\mu\mathrm{m}\). The stereo angle between the strip directions on the two
sides is \(7.5^\circ\), resulting in an effective two-dimensional pitch
of approximately \(58~\mu\mathrm{m}\times440~\mu\mathrm{m}\).

Signals from the strips are routed to the front-end boards through low-mass
aluminum--polyimide microcables~\cite{module_assembly} and read out with
STS/MUCH-XYTER (SMX) ASICs~\cite{smx}. The technological workflow of DSSD module assembly
and quality assurance is described in detail in Ref.~\cite{SheremetevThesis2025}. The module performance was validated
before installation in BM@N in proton-beam tests at the SC-1000
synchrocyclotron in Gatchina~\cite{DSSD_SC1000,DSSD_BMN}.

\section{Experimental conditions}
\label{sec:run9_operation_analysis}

\subsection{Integration of the VSP readout into the BM@N DAQ}
\label{subsec:daq_integration}

The VSP readout is based on the data-driven architecture derived from that
developed for the Silicon Tracking System~\cite{bmn_readout,STS_STATION}.
The front-end electronics uses SMX ASICs operating in a data-driven mode and
transmitting zero-suppressed hit frames. A hit is generated independently in
each readout channel when the signal amplitude exceeds the preset threshold of
the lower ADC comparator. Each VSP hit frame contains time-stamp and amplitude
information.

The SMX front-end operates continuously and does not receive an event trigger.
The selection of VSP hits is therefore performed in the backend electronics.
Incoming BM@N trigger signals are assigned time stamps, which are compared with
the time stamps of the continuous VSP hit stream. Hits falling within a
\(2~\mu\mathrm{s}\) time window around the trigger time were selected.

At the server level, the VSP readout software performs data processing, online
monitoring, and interaction with the common BM@N DAQ. The selected VSP data
are packed in the MPDRawData format~\cite{bmn_readout} and transferred to the
BM@N Event Builder.

\subsection{Data-taking conditions}
\label{subsec:vsp_data_taking_conditions}

Run~9 data taking lasted from 17 January to 4 April 2026. During this period,
the VSP station operated without critical failures affecting the mechanics,
cooling, or readout chain.

The sensors were biased at \(100~\mathrm{V}\), and the discriminator thresholds
were set to approximately \(6\times10^{3}\) electrons. Under these conditions,
the full-system noise-hit rate measured without beam in data-driven mode was
about \(1~\mathrm{kHz}\), corresponding to a probability of approximately
\(0.2\%\) for a random noise hit within a \(2~\mu\mathrm{s}\) trigger window.

Water cooling was used to stabilize the front-end electronics during
operation at temperatures up to \(40^{\circ}\mathrm{C}\). Owing to the low
sensor leakage currents and the sensor power dissipation of about
\(0.1~\mathrm{mW}\), no dedicated sensor cooling was required. The sensors
were cooled mainly by natural convection and heat exchange with the
water-cooled radiators. Under these conditions, the sensor temperature was
typically about \(25^{\circ}\mathrm{C}\), while the ambient temperature
varied from approximately \(25^{\circ}\mathrm{C}\) to
\(37^{\circ}\mathrm{C}\).

The performance results presented below are based on two \(3.0~A\mathrm{GeV}\) Xe+Sn datasets. A calibration dataset was recorded without
magnetic field and a physics dataset was recorded with the nominal BM@N magnetic
field.

\subsection{Data analysis}
\label{subsec:data_analysis}

The following coordinate and angular conventions are used throughout the
analysis. The \(z\) axis is directed along the beam, while \(x\) and \(y\) are
the transverse coordinates oriented perpendicular and parallel to the magnetic
field direction, respectively, as shown in
Fig.~\ref{fig:vsp_halfstation}(a). Track directions are parameterized by the
slopes \(t_x=dx/dz\) and \(t_y=dy/dz\).

The polar angle \(\theta\) is defined as the angle between the particle
direction and the beam axis \(z\). For detector-response studies, the track
incidence angle \(\alpha_s\) is defined separately for each sensor side in the
plane orthogonal to the corresponding strip direction. The two incidence
angles are denoted \(\alpha_{L0}\) for the n-side (L0) and
\(\alpha_{L1}\) for the p-side (L1).

\begin{figure}[htbp]
    \centering
    \includegraphics[width=0.88\textwidth]
    {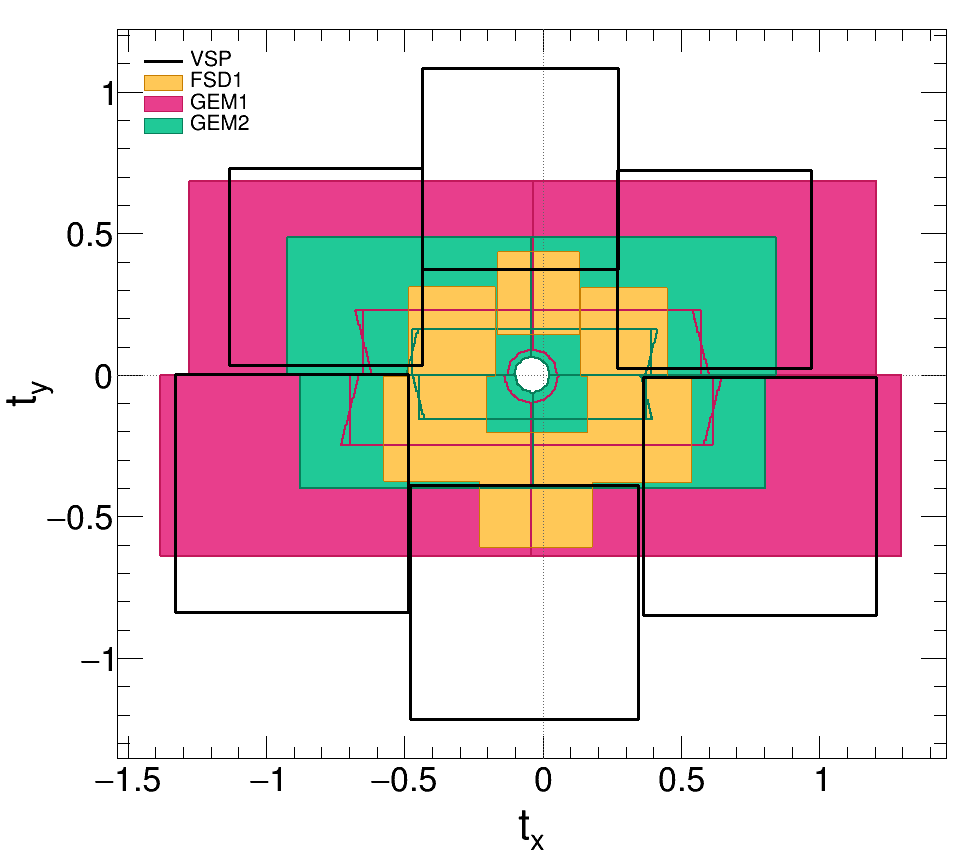}
    \caption{Angular acceptance of the VSP, FSD1, GEM1, and GEM2 in the
    \((t_x,t_y)\) plane.}
    \label{fig:tracking_acceptance}
\end{figure}

The angular acceptance of the VSP relative to the first FSD and GEM stations
is shown in Fig.~\ref{fig:tracking_acceptance}. The VSP has only a limited
geometrical overlap with the first FSD station, whereas its overlap with the
first GEM stations is substantially larger. Consequently, several
complementary tracking configurations were used for the detector-performance
study.

Raw data were decoded using the BmnRoot framework, which is based on
FairRoot~\cite{FairRoot}. CTS tracks were reconstructed with the regular
offline tracking procedure. Track candidates were built with the Vector Finder
toolkit~\cite{vector_finder} and fitted using a Kalman-filter
procedure~\cite{kalman_tracking}. These tracks were used for the track
impact-parameter study.

The primary vertex used in the analysis was reconstructed with the standard
BM@N Kalman-based vertex-fitting procedure~\cite{bmn_strange_reco}.

For alignment and geometrical cross-checks, a separate reference was used,
based on the target reference point reconstructed with the SiBT together with measurements in the first GEM station. This configuration
provides the maximum angular coverage. The alignment procedure and
the resulting constants are documented in BM@N internal
reports~\cite{RufanovAlignmentRun9,RufanovSiRotations2026}.

To extend track reconstruction towards midrapidity, short VSP--GEM tracklets
were also constructed using the three-dimensional primary-vertex point, one
VSP hit, and measurements in the first two GEM stations. These tracklets
exploit the large geometrical overlap between the VSP and the first GEM
stations and extend the tracklet acceptance beyond that of the CTS.
\section{Results}
\label{sec:results}

\subsection{Detector response and in situ performance}
\label{subsec:detector_response}

\subsubsection{Time resolution}
\label{subsubsec:time_resolution}

The VSP time resolution was evaluated from the distribution of VSP hit time
stamps relative to the time stamp assigned by the VSP readout to the trigger
pulse.

\begin{figure}[htbp]
\centering

\begin{minipage}[t]{0.49\textwidth}
    \centering
    \includegraphics[width=\linewidth]{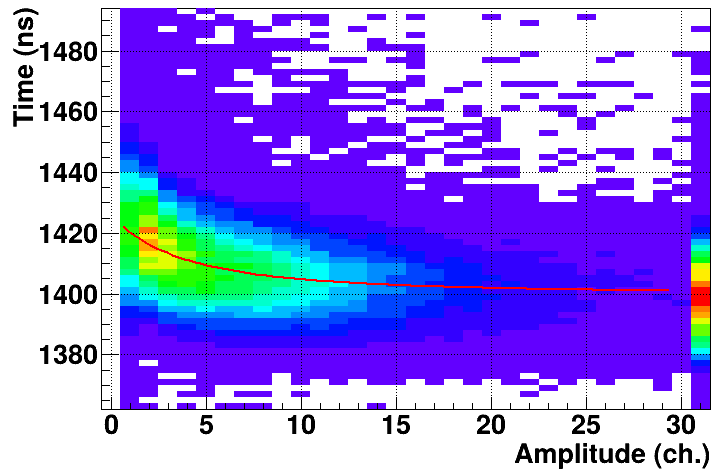}
\panelmark{a}
\end{minipage}
\hfill
\begin{minipage}[t]{0.49\textwidth}
    \centering
    \includegraphics[width=\linewidth]{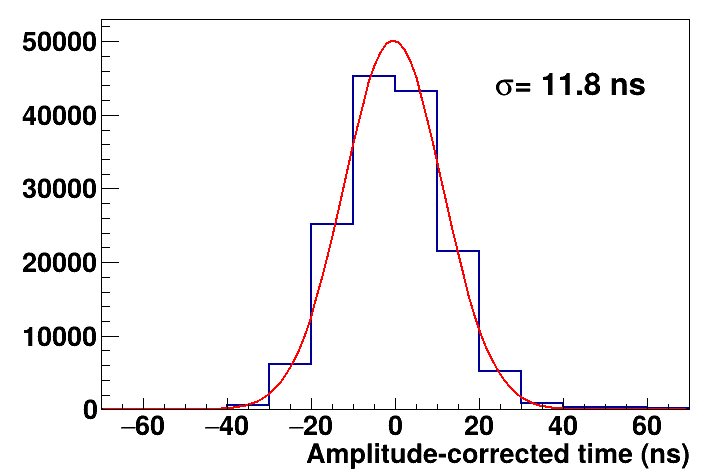}
\panelmark{b}
\end{minipage}

\caption{VSP hit-time response.
(a) Measured hit time as a function of signal amplitude before the time-slewing
correction. The red curve shows the fitted parameterization used to derive the
amplitude-dependent correction.
(b) Distribution of the corrected VSP hit time relative to the trigger pulse.
A Gaussian fit gives \(\sigma_t=11.8~\mathrm{ns}\).}
\label{fig:timing_corrected}
\end{figure}

The amplitude dependence of the hit time shown in
Fig.~\ref{fig:timing_corrected}(a) was fitted using the parameterization
described in Ref.~\cite{DSSD_BMN}; the fitted parameters were then used for the
time-slewing correction. After the correction, the hit--trigger time distribution exhibits
a well-defined peak, as shown in Fig.~\ref{fig:timing_corrected}(b). A Gaussian
fit gives a standard deviation of \(\sigma_t=11.8~\mathrm{ns}\). This value
represents the time resolution of the VSP relative to the trigger pulse and
includes contributions from the front-end time stamping, residual time-slewing
effects, trigger-reference jitter, and readout synchronization.

The observed time correlation demonstrates the successful matching of the
free-streaming VSP data with the BM@N trigger stream. The achieved 
time resolution is sufficient for selecting in-time VSP hits within the \(2~\mu\mathrm{s}\)
trigger window and suppressing out-of-time hits in the data-driven
readout.

\subsubsection{Amplitude measurement and signal-to-noise ratio}
\label{subsubsec:snr}

The cluster signal-to-noise ratio (SNR) was studied for VSP hits associated
with local SiBT-target--VSP--GEM1 track candidates in the calibration dataset
recorded without magnetic field.  Cluster amplitudes were converted to charge using module- and side-dependent calibration constants. The calibration procedure for the
SMX-based silicon readout is described in
Ref.~\cite{Shitenkov_thesis}. Measurements of the internal calibration
circuits for different SMX chips showed an RMS variation of the calibration
scale of about \(4\%\).

\begin{figure}[t]
  \centering
  \includegraphics[width=0.98\linewidth]
  {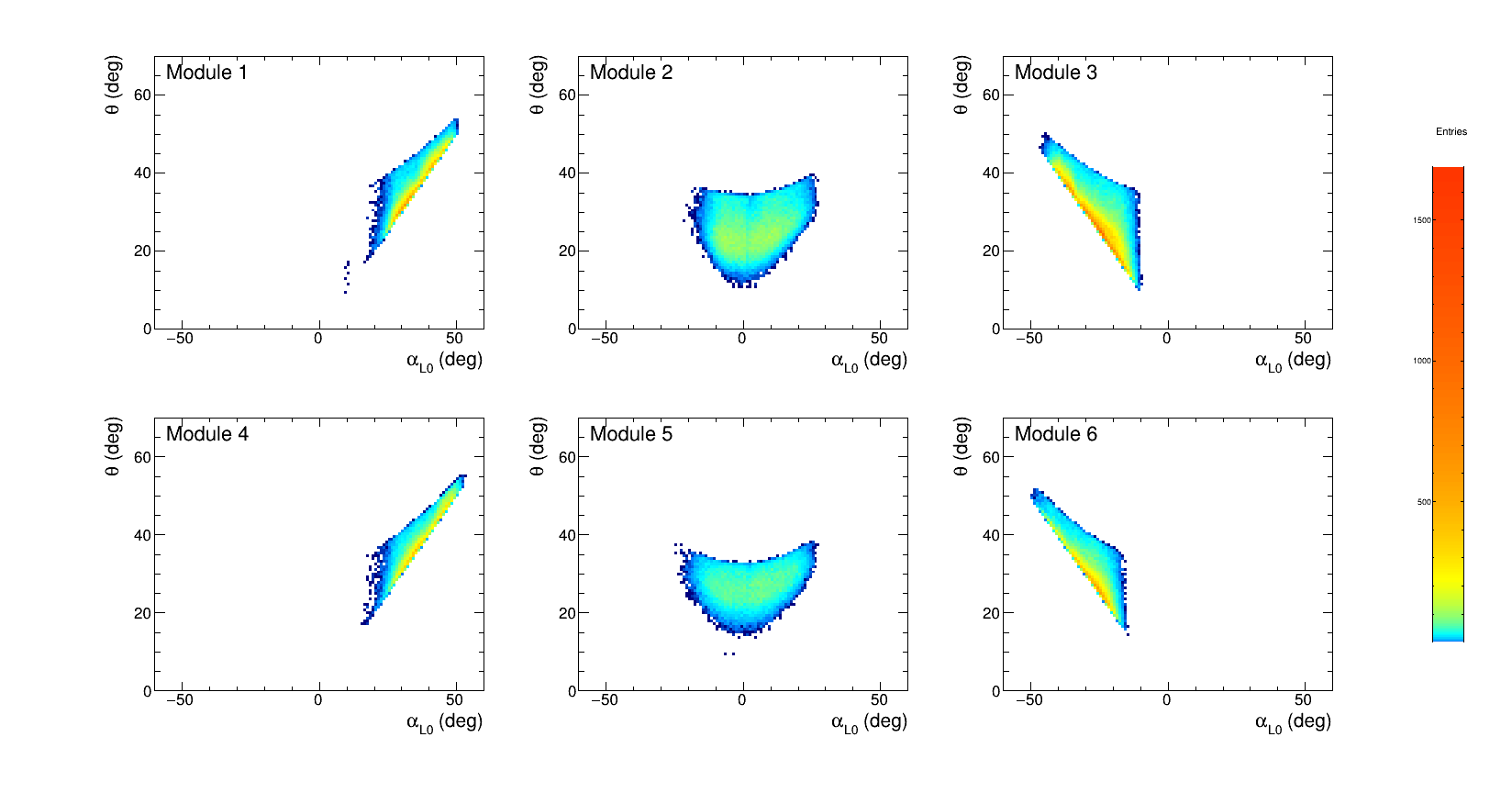}
  \caption{Module-resolved angular coverage of reconstructed tracks in the
  \(\theta\)--\(\alpha_{L0}\) plane. The angular definitions are given in
  Sec.~\ref{subsec:data_analysis}.}
  \label{fig:vsp_theta_alpha}
\end{figure}

The six VSP modules cover different regions of the angular phase space.
The definitions of the polar angle \(\theta\) and the track incidence angle
\(\alpha_{L0}\) are given in Sec.~\ref{subsec:data_analysis}. Their
module-resolved coverage in the \(\theta\)--\(\alpha_{L0}\) plane is shown in
Fig.~\ref{fig:vsp_theta_alpha}.

The two angles affect the detector response in different ways. The polar angle
\(\theta\) determines the particle path length in the silicon, whereas the
track incidence angle \(\alpha_{L0}\) affects charge sharing across the
strips. The different angular coverage of the modules therefore contributes
to the observed module dependence of the cluster response.

The cluster signal-to-noise ratio was calculated as
\begin{equation}
  \mathrm{SNR} =
  \frac{S_{\mathrm{cluster}}}
       {N_{m,s}\sqrt{n_{strips}}},
  \label{eq:vsp_snr}
\end{equation}
where \(S_{\mathrm{cluster}}\) is the sum of the hit-induced charges of the
strips forming the cluster, \(n_{strips}\) is the number of strips contributing to the
cluster, and \(N_{m,s}\) is the mean single-strip noise for module \(m\)
and sensor side \(s\). Clusters in the double-metallization region of the
p-side (L1) were excluded from the analysis because of the higher strip noise
observed in this region, as discussed in Ref.~\cite{DSSD_BMN}.

\begin{figure}[t]
  \centering
  \includegraphics[width=0.95\linewidth]
  {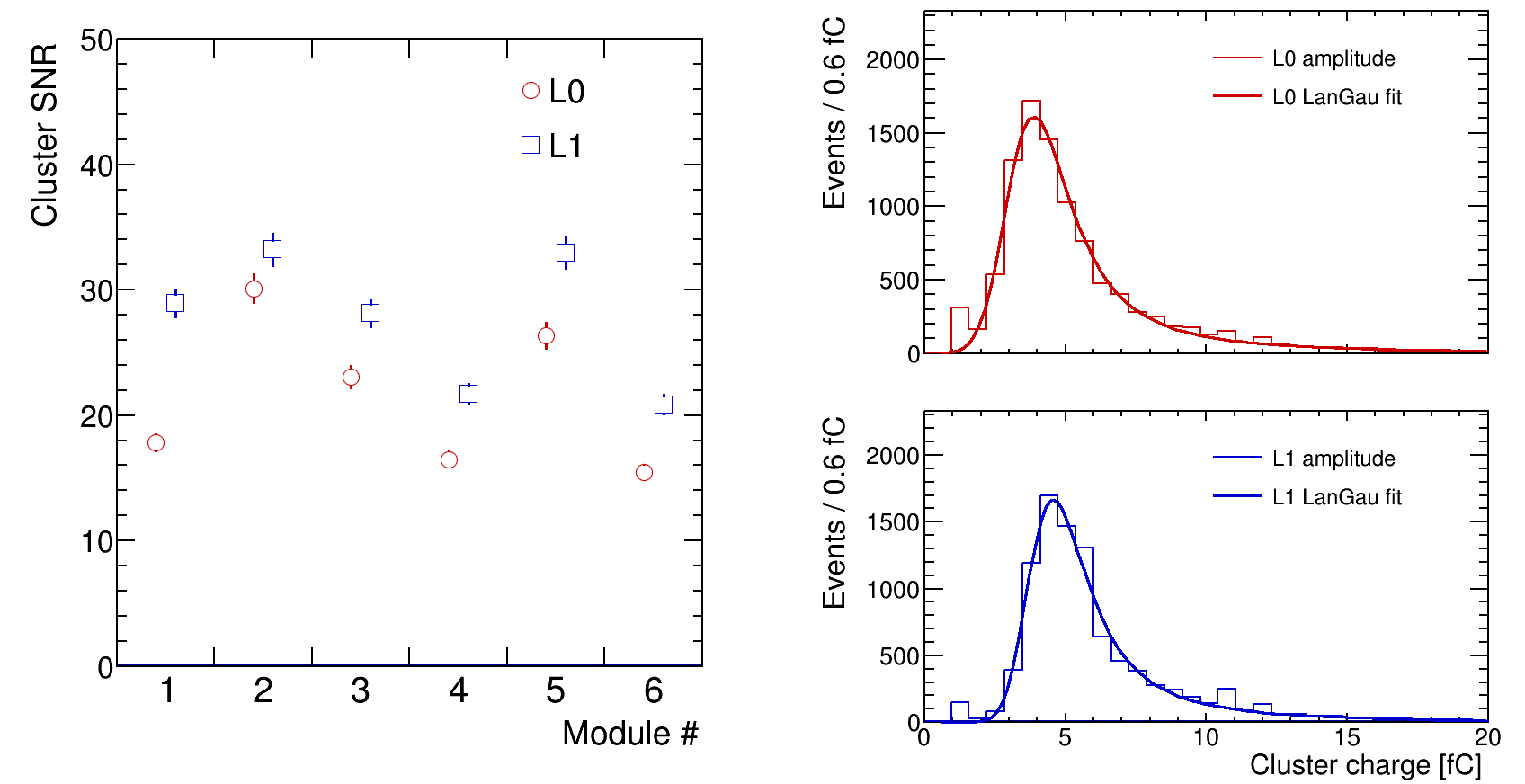}
  \caption{Cluster charge and signal-to-noise response of the VSP.
  Left: most probable cluster SNR for the six VSP modules and two sensor
  sides, obtained from Landau--Gaussian convolution fits. The error bars
  represent the statistical uncertainties of the fitted MPV values.
  Right: representative calibrated cluster-charge distributions for
  module~5 with Landau--Gaussian convolution fits.}
  \label{fig:vsp_snr_charge}
\end{figure}

Figure~\ref{fig:vsp_snr_charge} summarizes the measured cluster response.
The left panel shows the most probable value (MPV) of the SNR distribution
for each module and sensor side, obtained from Landau--Gaussian convolution
fits. The error bars represent the statistical uncertainties of the fitted
MPV values. The right panel shows representative calibrated cluster-charge
distributions for module~5.

The fitted SNR MPVs range from \(15.5\) to \(30.1\) on the n-side (L0) and
from \(20.9\) to \(33.2\) on the p-side (L1). The module-to-module variation
is consistent with the different angular coverage shown in
Fig.~\ref{fig:vsp_theta_alpha}. Modules~2 and~5 are dominated by compact
one- and two-strip clusters, whereas larger clusters are more frequent in
modules covering broader ranges of track incidence angles. Since the cluster
noise increases with the number of strips contributing to the cluster, the
different cluster-size composition contributes to the observed SNR variation.

\subsubsection{Lorentz drift}
\label{subsubsec:lorentz_drift}

For the analysis of data recorded with the nominal magnetic field, the
Lorentz drift of charge carriers in the silicon sensors has to be taken into
account, since it introduces an additional displacement of the reconstructed
hit position. The size of this effect was therefore evaluated directly from
the Run~9 data by comparing the detector response in the physics dataset
recorded with the nominal magnetic field of \(0.5~\mathrm{T}\) with that in
the calibration dataset recorded without magnetic field.

The induced signal is described by the Shockley--Ramo theorem
~\cite{Shockley1938}, according to which the current induced on a given
electrode is determined by the carrier motion in the corresponding weighting
field. Although both electrons and holes contribute to the total induced
signal, the weighting field is concentrated near the collecting strips.
Consequently, the n-side (L0) signal is dominated by electrons approaching
the \(n^+\) strips, whereas the p-side (L1) signal is dominated by holes
approaching the \(p^+\) strips.

A magnetic field deflects the carrier drift direction from the electric-field
direction. The Lorentz angle is given by
\begin{equation}
  \tan\theta_L =
  \mu_H B
  =
  r_H\mu_d(E,T)B,
  \label{eq:lorentz_angle}
\end{equation}
where \(\mu_d\) is the drift mobility, \(r_H\) is the Hall scattering factor,
and \(B\) denotes the magnetic-field component perpendicular to the sensor
electric field. Electron and hole deflections differ because of their
different Hall mobilities.

\begin{figure}[htbp]
\centering
\begin{minipage}{0.48\textwidth}
\centering
\includegraphics[width=\textwidth]{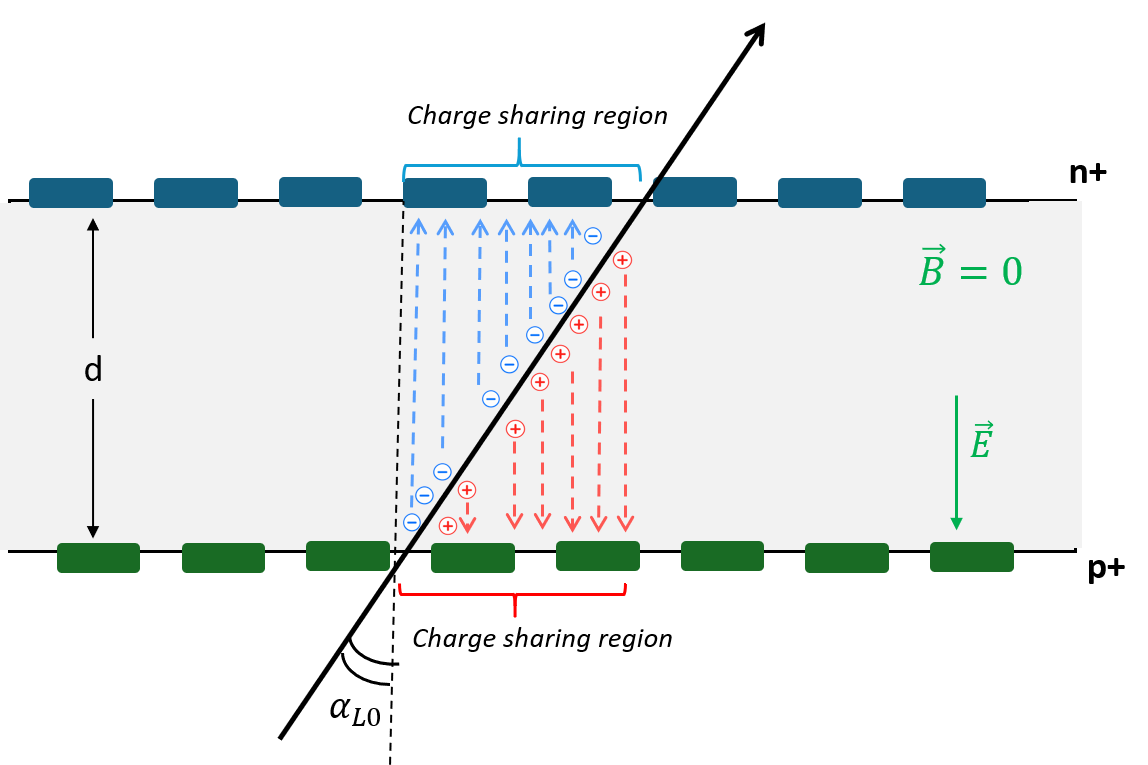}\\[-1mm]
\panelmark{a} 
\end{minipage}
\hfill
\begin{minipage}{0.48\textwidth}
\centering
\includegraphics[width=\textwidth]{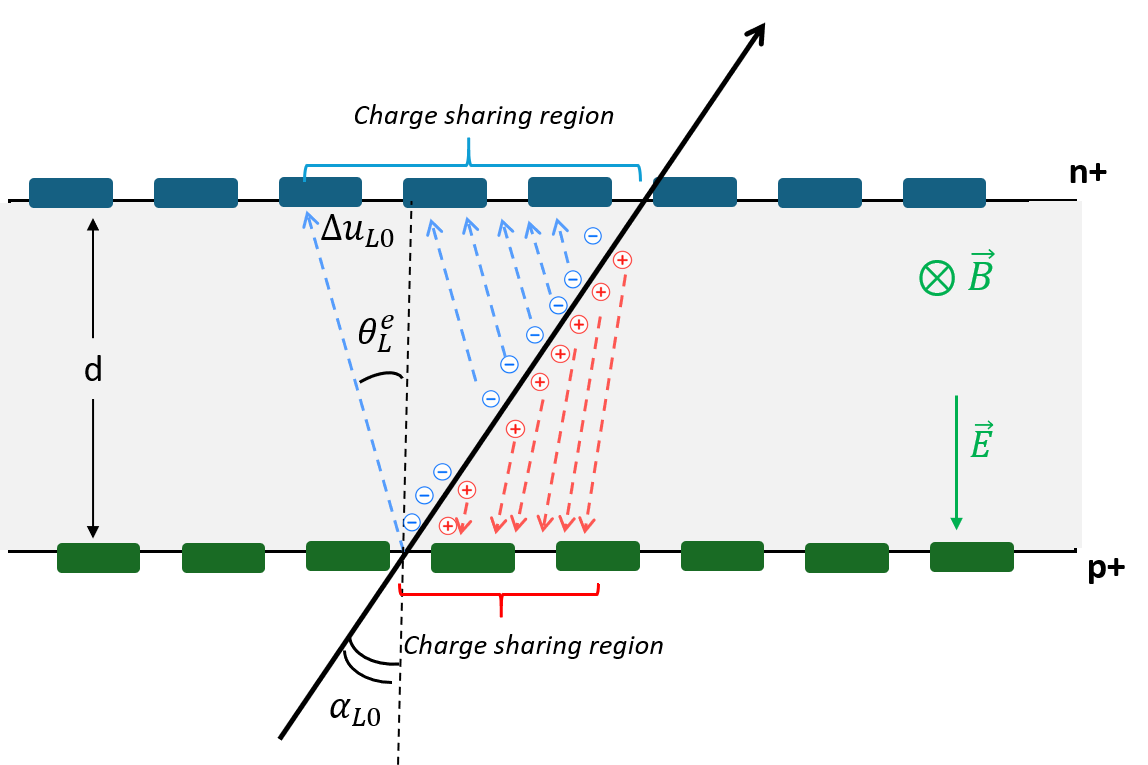}\\[-1mm]
\panelmark{b} 
\end{minipage}

\caption{Schematic view of charge-carrier drift in a double-sided silicon strip
sensor without (a) and with (b) magnetic field. It does not
include diffusion, strip-weighting-field effects, capacitive charge sharing or
threshold effects. The scheme is shown for L0 in
the $X$--$Z$ projection; the L1 case is treated analogously in the $U$--$Z$
projection.}
\label{fig:vsp_lorentz_schematic} 
\end{figure}

Figure~\ref{fig:vsp_lorentz_schematic} shows the L0 case in the $X$--$Z$
projection. For L1, the projected incidence angle $\alpha_{L1}$ is defined
analogously in the local $U$--$Z$ projection. The different L0 and L1 shifts
follow from Eq.~\eqref{eq:lorentz_angle}, because the electron and hole
Lorentz angles, $\theta_L^e$ and $\theta_L^h$, are different.

For approximately uniform ionization through a sensor of thickness \(d\),
the charge carriers are produced uniformly along the drift direction.
For a carrier with drift distance \(z\) to the collecting electrode, the
Lorentz-induced transverse displacement is \(z\tan\theta_L\). Averaging over
the sensor thickness gives the corresponding displacement of the charge
centroid,
\begin{equation}
  \left|\Delta u_{\mathrm{CoG}}\right|
  \simeq
  \frac{1}{d}\int_{0}^{d} z\tan|\theta_L|\,dz
  =
  \frac{d}{2}\tan|\theta_L|.
  \label{eq:lorentz_cog_shift}
\end{equation}

The absolute cluster size depends on several detector-response effects,
including diffusion, non-uniform weighting fields near the strips, capacitive
charge sharing and thresholds. The observable used here is the displacement of
the minimum of the measured \(\langle N_{\mathrm{strips}}\rangle(\alpha_s)\)
curve between the zero-field and magnetic-field data for the same sensor side.

The minimum occurs when the geometrical charge sharing is compensated by the
Lorentz deflection of the charge carriers~\cite{Campabadal2005}. This makes the method less sensitive to alignment offsets than a direct
comparison of mean track--hit residuals.

\begin{figure}[htbp]
\centering
\begingroup
\setlength{\fboxsep}{0pt}
\begin{overpic}[width=0.92\linewidth]{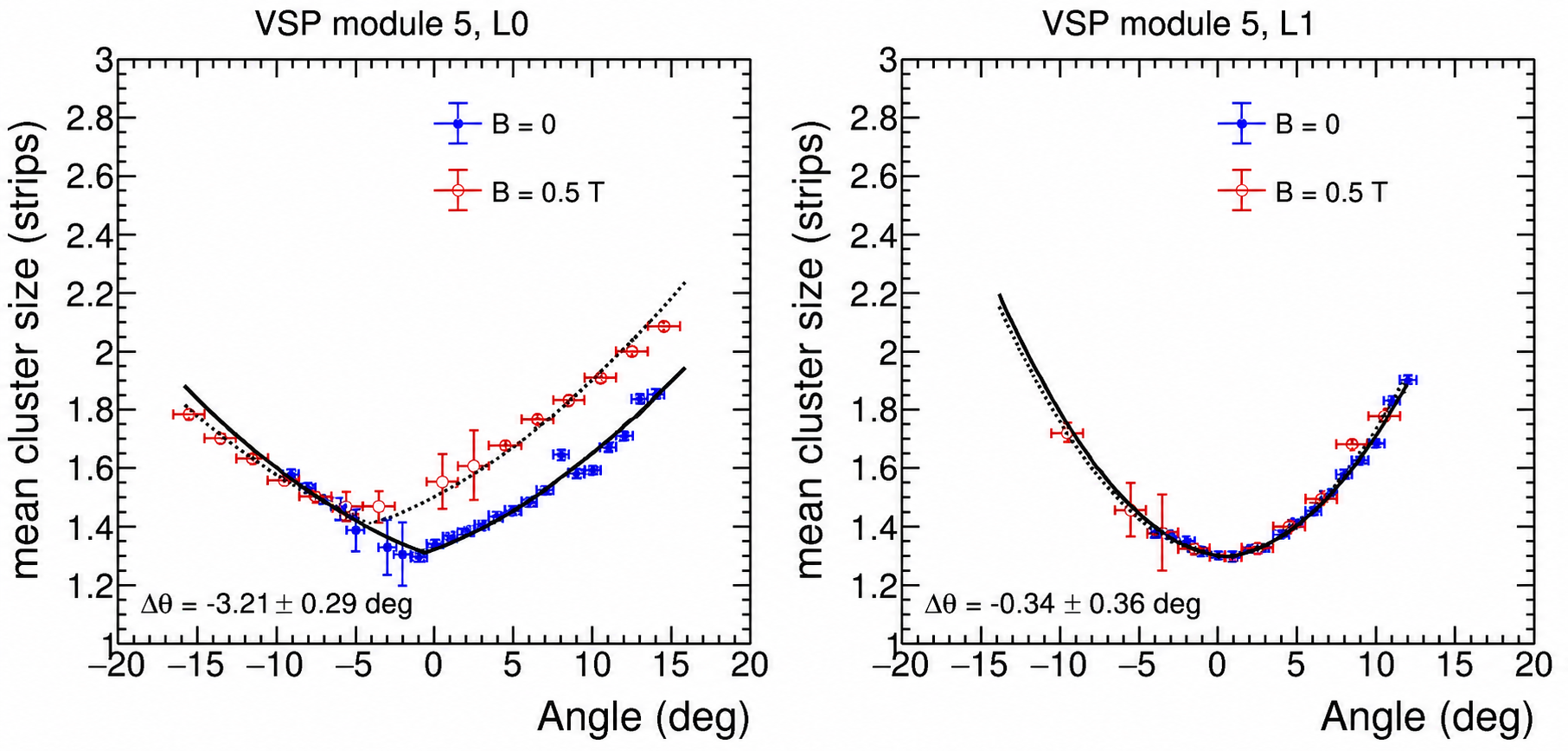}
  \put(25.0,0.0){\colorbox{white}{\makebox(24.0,3.8){\small $\alpha_{L0}$ (deg)}}}
  \put(75.0,0.0){\colorbox{white}{\makebox(24.8,3.8){\small $\alpha_{L1}$ (deg)}}}
  \put(8.4,7.5){\colorbox{white}{\makebox(37.5,2.6)[l]{\fontsize{9}{11}\selectfont $\Delta\alpha_{L0}=(-3.2\pm0.3)^{\circ}$}}}
  \put(58.4,7.5){\colorbox{white}{\makebox(37.2,2.6)[l]{\fontsize{9}{11}\selectfont $\Delta\alpha_{L1}=(-0.3\pm0.4)^{\circ}$}}}
\end{overpic}
\endgroup
\caption{Mean cluster size as a function of the track incidence angle for the
n-side (L0) and p-side (L1) of VSP module~5. The calibration dataset recorded
without magnetic field is compared with the physics dataset recorded with the
nominal magnetic field. The curves show parabolic fits around the minima.}
\label{fig:vsp_cluster_size_shift}
\end{figure}

Modules~5 and~2 were selected for this study because the region around the
minimum is well populated for both modules, as shown in
Fig.~\ref{fig:vsp_theta_alpha}. 
The calibration and physics datasets were fitted separately with
\begin{equation}
  \left\langle N_{\mathrm{strips}}\right\rangle(\alpha_s) =
  N_{\min}+a_s(\alpha_s-\alpha_{\min,s})^2,
  \label{eq:cluster_size_angle_fit}
\end{equation}
where \(\alpha_{\min,s}\) denotes the fitted minimum. Using the fitted values
\(\alpha_{\min,s}^{0}\) and \(\alpha_{\min,s}^{B}\) for the calibration and
physics datasets, respectively, the field-induced shift is defined as
\[
\Delta\alpha_s=
\alpha_{\min,s}^{B}-\alpha_{\min,s}^{0},
\qquad s=L0,L1.
\]
The module~5 results are shown in
Fig.~\ref{fig:vsp_cluster_size_shift}.

The measured shifts are
\(\Delta\alpha_{L0}=(-3.2\pm0.3)^{\circ}\) and
\(\Delta\alpha_{L1}=(-0.3\pm0.4)^{\circ}\). Interpreting the shifts of the
cluster-size minima as Lorentz-angle estimates,
\(|\theta_L|\simeq|\Delta\alpha_s|\), Eq.~\ref{eq:lorentz_cog_shift} gives
\[
|\Delta u_{\mathrm{CoG},L0}|=(9.0\pm0.8)~\mu\mathrm{m},
\qquad
|\Delta u_{\mathrm{CoG},L1}|=(0.8\pm1.1)~\mu\mathrm{m}.
\]
Applying the same procedure to module~2 gives
\(\Delta\alpha_{L0}=(-3.7\pm0.4)^{\circ}\), consistent with the result for
module~5.

For all modules, the magnetic-field component perpendicular to the sensor electric
field is \(B\simeq0.5~\mathrm{T}\). For a sensor temperature of
\(25^{\circ}\mathrm{C}\), a nominal bias voltage of \(100~\mathrm{V}\), and
a sensor thickness of \(d=320~\mu\mathrm{m}\), the mobility model of
Ref.~\cite{Lari2001} gives \(\theta_L^{e}=3.7^{\circ}\) and
\(\theta_L^{h}=0.9^{\circ}\). Equation~\ref{eq:lorentz_cog_shift} then gives
expected charge-centroid shifts of \(10.5~\mu\mathrm{m}\) for electrons and
\(2.4~\mu\mathrm{m}\) for holes.

The measured n-side (L0) result agrees with the predicted electron Lorentz
angle. The smaller p-side (L1) shift is consistent with the lower hole Hall
mobility and is also compatible with zero within the present uncertainty.
\subsection{Tracking performance}
\label{subsec:tracking_performance}

\subsubsection{Track--hit residuals}
\label{subsubsec:track_hit_residuals}

Track--hit residuals were measured in the calibration dataset recorded
without magnetic field using two independent linear reference configurations,
one based on the FSD stations and one on the GEM stations. The VSP hit was
excluded from both reference-track fits.

The residuals are defined as
\(\Delta X=X_{\mathrm{VSP}}-X_{\mathrm{pred}}\) and
\(\Delta Y=Y_{\mathrm{VSP}}-Y_{\mathrm{pred}}\), where
\(X_{\mathrm{pred}}\) and \(Y_{\mathrm{pred}}\) are the reference-track
coordinates extrapolated to the VSP plane. Reference tracks were required to
satisfy \(\chi^2/\mathrm{ndf}<1.5\).

\begin{figure}[htbp]
\centering
\begin{minipage}[t]{0.49\textwidth}
  \centering
  \includegraphics[width=\linewidth]{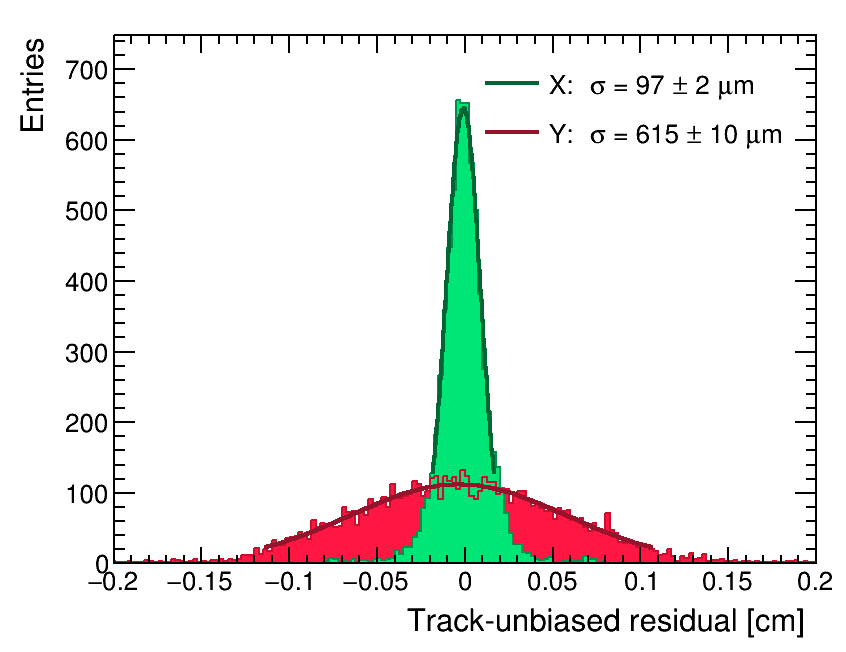}
  \panelmark{a}
\end{minipage}
\hfill
\begin{minipage}[t]{0.49\textwidth}
  \centering
  \includegraphics[width=\linewidth]{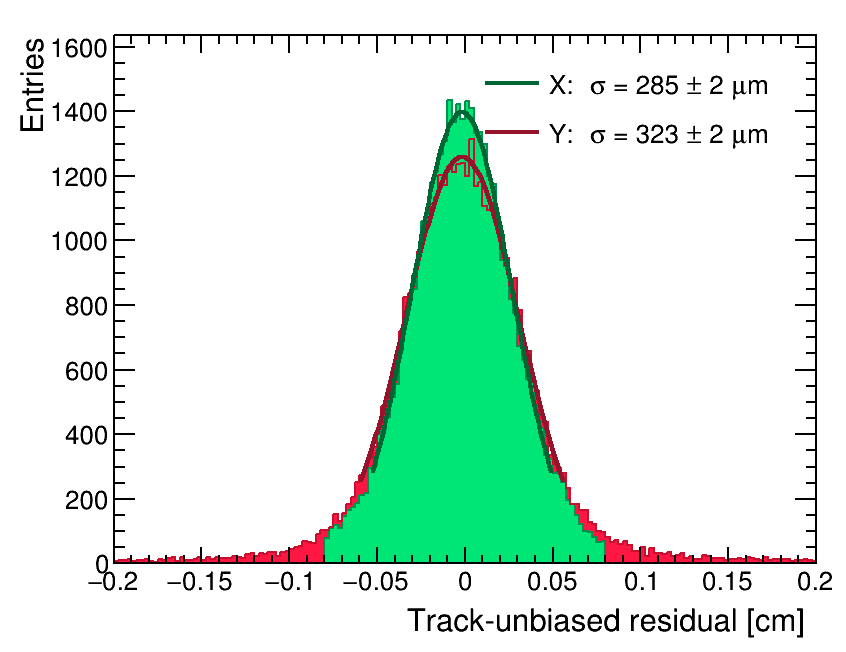}
  \panelmark{b}
\end{minipage}
\caption{Unbiased track--hit residuals measured in the calibration dataset
recorded without magnetic field.
(a) Linear reference track based on the FSD stations.
(b) Linear reference track based on the GEM stations.
The VSP hit is excluded from both reference fits. The solid curves show
Gaussian fits to the central region of the distributions.}
\label{fig:vsp_unbiased_residuals_refs}
\end{figure}

Figure~\ref{fig:vsp_unbiased_residuals_refs} shows the resulting unbiased
track--hit residuals. The standard deviations of Gaussian fits to the central
part of the distributions are
\(\sigma_X=(97\pm2)~\mu\mathrm{m}\) and
\(\sigma_Y=(615\pm10)~\mu\mathrm{m}\) for the FSD reference, and
\(\sigma_X=(285\pm2)~\mu\mathrm{m}\) and
\(\sigma_Y=(323\pm2)~\mu\mathrm{m}\) for the GEM reference.

The two reference configurations have different extrapolation precisions. The
FSD reference provides better precision in \(X\), whereas the GEM reference
gives more similar resolutions in the two coordinates. The measured residual
widths are dominated by the reference-track extrapolation uncertainty and
therefore do not represent the intrinsic spatial resolution of the VSP.

A head--tail cluster-position estimator was also tested as an alternative to
the standard center-of-gravity (CoG) method for large track incidence
angles~\cite{Campabadal2005}. A small improvement of about
\(1~\mu\mathrm{m}\) was observed, which is negligible compared with the
residual widths and is consistent with their being dominated by the
reference-track extrapolation uncertainty.

\subsubsection{Hit-matching efficiency}
\label{subsubsec:hit_reconstruction_efficiency}

The VSP hit-matching efficiency was measured in the calibration dataset
recorded without magnetic field using the two independent FSD- and GEM-based
reference configurations described in
Sec.~\ref{subsubsec:track_hit_residuals}. A compatible VSP hit was selected using
the normalized track--hit distance
\begin{equation}
 d_R =
 \sqrt{
 \left(\frac{\Delta X}{\sigma_X}\right)^2+
 \left(\frac{\Delta Y}{\sigma_Y}\right)^2},
 \label{eq:vsp_dR}
\end{equation}
where the reference-specific values of \(\sigma_X\) and \(\sigma_Y\) were
obtained from the residual distributions described in
Sec.~\ref{subsubsec:track_hit_residuals}. The nominal matching
requirement was \(d_R<5\).

To suppress edge effects, only reference tracks crossing the VSP sensitive
area at least \(5~\mathrm{mm}\) from its edges were retained. The
hit-matching efficiency was defined as
\(\varepsilon=N_{\mathrm{matched}}/N_{\mathrm{tracks}}\), where
\(N_{\mathrm{tracks}}\) is the number of independently reconstructed reference
tracks crossing the selected fiducial region and \(N_{\mathrm{matched}}\) is
the number of those tracks with a compatible VSP hit.

Because of the relatively high VSP hit occupancy, accidental matches between
reference tracks and VSP hits can occur. Their contribution was estimated
using event mixing. The same reference tracks were matched to VSP hits taken
from different events using the same geometrical and matching criteria. The
corrected efficiency was calculated as
\begin{equation}
  \varepsilon_{\mathrm{corr}} =
  \frac{\varepsilon_{\mathrm{same}}-\varepsilon_{\mathrm{mix}}}
       {1-\varepsilon_{\mathrm{mix}}},
  \label{eq:vsp_eff_corr}
\end{equation}
where \(\varepsilon_{\mathrm{same}}\) and
\(\varepsilon_{\mathrm{mix}}\) are the matching probabilities for the
same-event and mixed-event samples, respectively. The statistical uncertainty
on \(\varepsilon_{\mathrm{corr}}\) was obtained by propagating their binomial
uncertainties.

\begin{figure}[htbp]
\centering
\begin{minipage}[t]{0.49\textwidth}
  \centering
  \includegraphics[width=\linewidth]{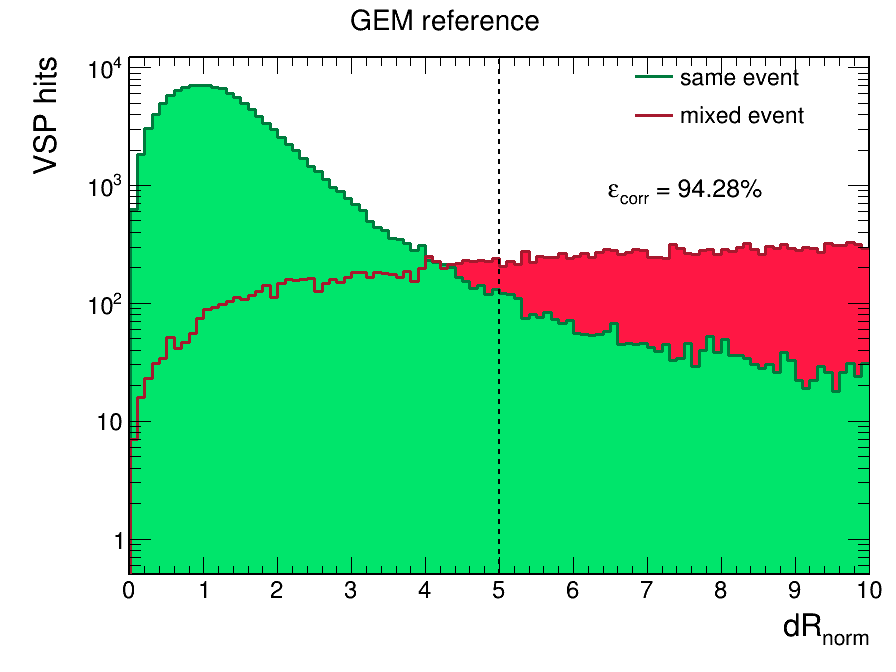}
  \panelmark{a}
\end{minipage}
\hfill
\begin{minipage}[t]{0.49\textwidth}
  \centering
  \includegraphics[width=\linewidth]{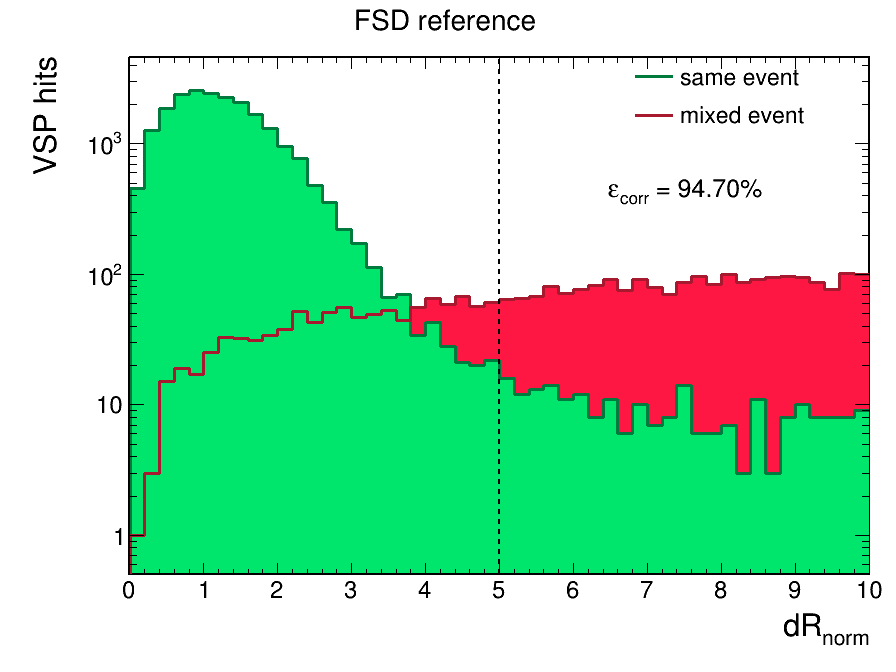}
  \panelmark{b}
\end{minipage}
\caption{Normalized VSP track--hit matching distance for
(a) the GEM reference and (b) the FSD reference. Green distributions
correspond to VSP hits from the same event, while red distributions show the
mixed-event control sample used to estimate accidental track matching. The
dashed line indicates the nominal requirement \(d_R<5\).}
\label{fig:vsp_efficiency_dr}
\end{figure}

Figure~\ref{fig:vsp_efficiency_dr} shows the same-event and mixed-event
\(d_R\) distributions. Both reference configurations show a clear excess of
same-event matches at small \(d_R\).

\begin{table}[htbp]
\centering
\caption{Integrated VSP hit-matching efficiency obtained with the two
independent reference configurations. The uncertainty quoted for
\(\varepsilon_{\mathrm{corr}}\) is statistical.}
\label{tab:vsp_efficiency_summary}

\begin{tabular}{lccc}
\toprule
Reference &
\begin{tabular}{c}
\(\varepsilon_{\mathrm{same}}\),\\
\%
\end{tabular} &
\begin{tabular}{c}
\(\varepsilon_{\mathrm{mix}}\),\\
\%
\end{tabular} &
\begin{tabular}{c}
\(\varepsilon_{\mathrm{corr}}\),\\
\%
\end{tabular}
\\
\midrule

GEM reference & 94.6 & 5.5 & \(94.3\pm0.1\) \\
FSD reference & 94.9 & 4.4 & \(94.7\pm0.2\) \\

\bottomrule
\end{tabular}
\end{table}
As summarized in Table~\ref{tab:vsp_efficiency_summary}, the two independent
reference configurations give consistent corrected VSP hit-matching
efficiencies of about \(95\%\) in the selected fiducial region. For comparison,
dedicated proton-beam tests of the same DSSD technology demonstrated detection
efficiencies above \(99\%\) in fully operational sensor
regions~\cite{DSSD_BMN}.

\subsection{Contribution of the VSP to BM@N tracking}
\label{subsec:tracking_impact}

\subsubsection{Track impact parameter}
\label{subsubsec:track_impact_parameter}

Precise determination of decay vertices is crucial for the reconstruction of
complex decay topologies. The contribution of the near-target VSP measurement
to the vertexing performance was evaluated using the track impact parameter,
defined as the transverse distance of closest approach of the particle
trajectory to the primary vertex. In the present analysis, its \(x\) and \(y\)
projections were studied as a function of the inverse rigidity \(q/p\).

Two samples of CTS tracks were compared. \emph{VSP-start tracks}
have their first reconstructed tracking point in the VSP, whereas
\emph{FSD1-start tracks} begin in FSD1, the first tracking station downstream
of the VSP. Tracks were required to be associated with the reconstructed
primary vertex and to satisfy \(\theta>25^\circ\).

\begin{figure}[htbp]
\centering
\includegraphics[width=0.96\textwidth]{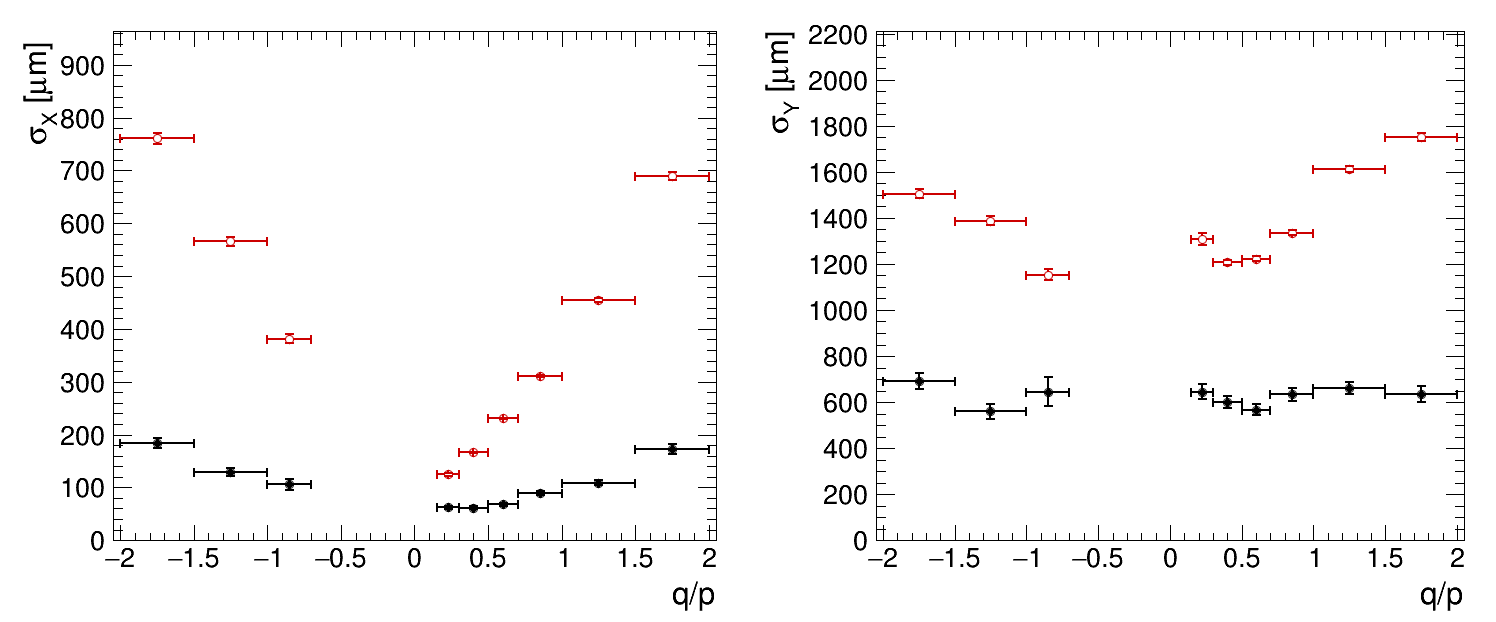}
\caption{Half-width \(\sigma\) of the central 68\% interval of the
\(x\)- and \(y\)-projected track impact-parameter distributions as a function
of the inverse rigidity \(q/p\). Solid (black) markers show VSP-start tracks, while
open (red) markers show FSD1-start tracks. The left and right panels correspond
to the \(x\) and \(y\) projections, respectively.}
\label{fig:vsp_impact_parameter_qp}
\end{figure}

The width of the impact-parameter distributions was characterized by
\(\sigma\), defined as the half-width of the central 68\% interval of the
corresponding distribution. The result is shown in
Fig.~\ref{fig:vsp_impact_parameter_qp}.

In all cases VSP-start tracks show narrower impact-parameter distributions over the studied
\(q/p\) range. The difference is most pronounced in the \(x\) (bending)
projection, where the widths for the VSP-start sample are approximately
\(50\text{--}200~\mu\mathrm{m}\). Smaller widths are also observed in the
\(y\) projection.

\subsubsection{Rigidity reconstruction with VSP--GEM tracklets}
\label{subsubsec:rigidity}

The large geometrical overlap of the VSP with the first GEM stations makes it
possible to construct short VSP--GEM tracklets in a region where the acceptance
of the CTS is reduced. This topology is of particular interest for extending
the reconstruction of high-$p_T$ protons towards midrapidity.

The potential extension of the tracking acceptance was first evaluated using
DCM--SMM Monte Carlo simulations of minimum-bias Xe+Sn collisions at
3~AGeV in the BM@N Run~9 geometry~\cite{DCMSMM}. A dedicated
reconstruction-level study of this VSP-assisted topology, including
magnetic-field propagation and combinatorial-background suppression, is
presented in Ref.~\cite{VSP_highpt}.

The phase space of reconstructible primary protons was compared for CTS tracks
and VSP--GEM tracklets. A proton was considered reconstructible by the CTS if
MC points were present in at least four distinct FSD or GEM stations. For the
VSP--GEM topology, reconstructibility required MC points in the VSP, GEM1, and
GEM2 together with the three-dimensional primary-vertex point. Since this
study is based on MC information rather than reconstructed detector hits, it
represents geometrical reconstructability rather than tracking efficiency.

\begin{figure}[htbp]
\centering
\includegraphics[width=0.98\textwidth]{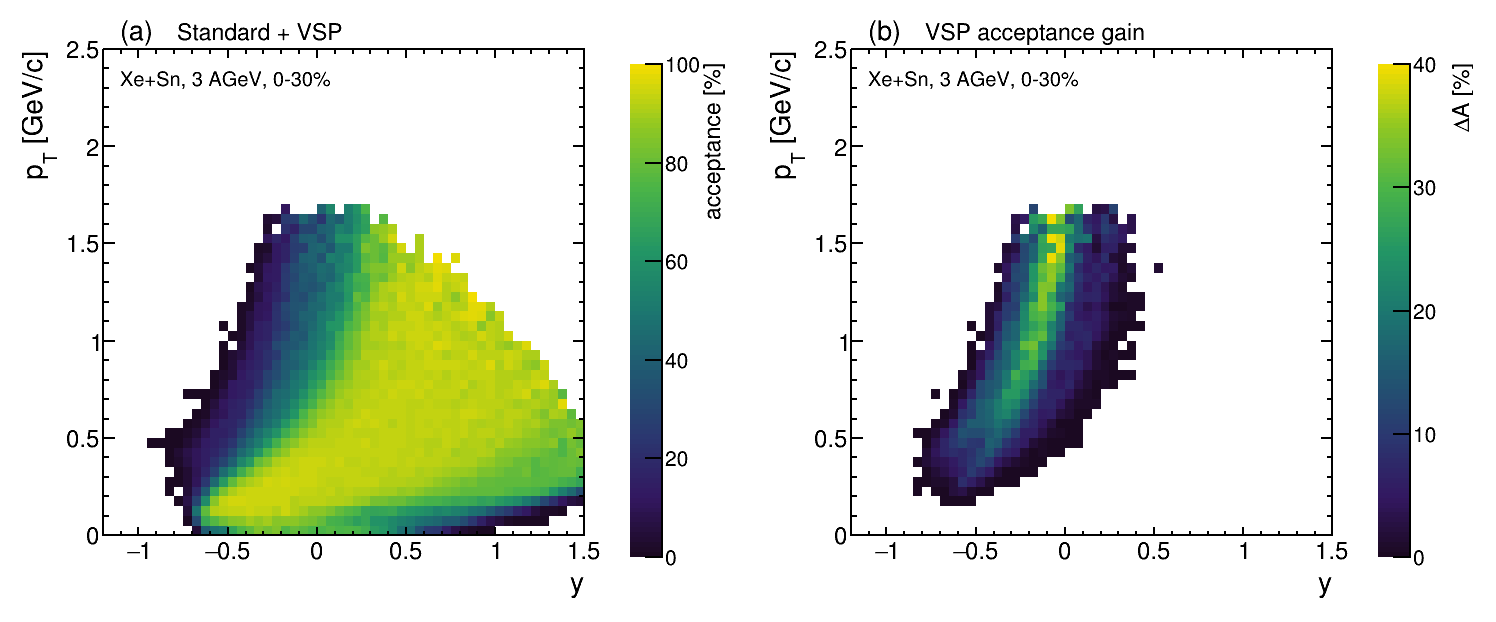}
\caption{Monte Carlo geometrical reconstructability of primary protons in
Xe+Sn collisions at \(3~A\mathrm{GeV}\) for the 0--30\% centrality class in
the \((y,p_T)\) plane, where \(y\) is the proton rapidity relative to the
center-of-mass rapidity, \(y=y_{\rm lab}-y_{\rm cm}\).
(a) Acceptance of the combined CTS and VSP--GEM tracklet configurations.
(b) Absolute increase
\(\Delta A=A_{\rm CTS+VSP\text{-}GEM}-A_{\rm CTS}\)
provided by the VSP--GEM tracklets.
The acceptance in each bin is defined as the fraction of generated primary
protons satisfying the corresponding MC-point reconstructability requirement.}
\label{fig:vsp_tracklet_acceptance}
\end{figure}

Figure~\ref{fig:vsp_tracklet_acceptance} shows the resulting acceptance for the
0--30\% centrality class in the \((y,p_T)\) plane, where \(y\) denotes the
proton rapidity relative to the center-of-mass rapidity,
\(y=y_{\rm lab}-y_{\rm cm}\); the proton mass is used in the rapidity
calculation. The VSP--GEM tracklet topology extends the
reconstructible phase space towards midrapidity and higher transverse
momenta.  For primary protons with
$p_T>1.2~\mathrm{GeV}/c$, the geometrical reconstructability increases
from 34.7\% for the CTS alone to 43.1\% when the VSP--GEM tracklet topology
is included. This corresponds to an absolute gain of about 8.5 percentage
points, or a relative increase of about 24\% with respect to the CTS.

Motivated by this geometrical acceptance gain, VSP--GEM tracklets were
reconstructed in the physics dataset recorded with the nominal magnetic field.
The tracklets were built using the three-dimensional primary-vertex point,
one VSP hit, and
measurements in the first two GEM stations. The inverse rigidity \(q/p\) was
determined using the BM@N magnetic-field map.

\begin{figure}[htbp]
\centering

\begin{minipage}[t]{0.49\textwidth}
\centering
\includegraphics[width=\linewidth]{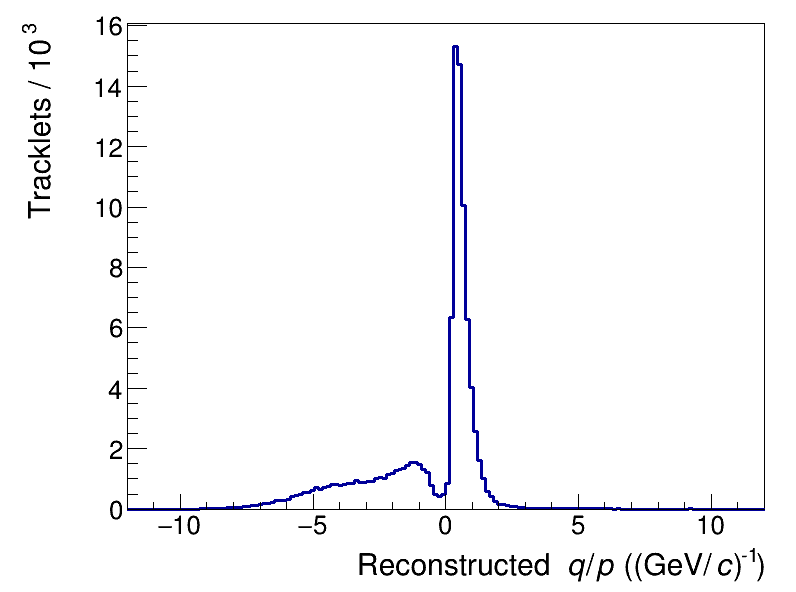}
\panelmark{a}
\end{minipage}
\hfill
\begin{minipage}[t]{0.49\textwidth}
\centering
\includegraphics[width=\linewidth]{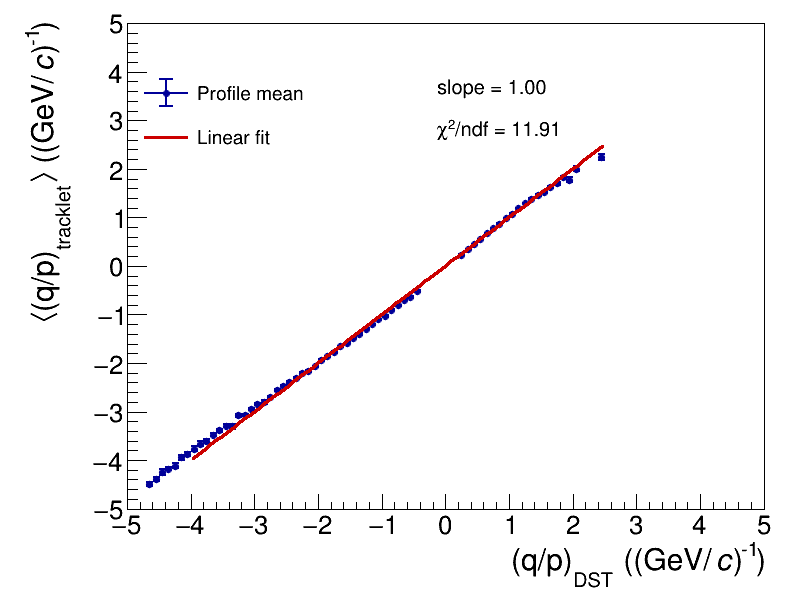}
\panelmark{b}
\end{minipage}

\caption{Rigidity reconstruction with VSP--GEM tracklets in the physics
dataset recorded with the nominal magnetic field.
(a) Distribution of the reconstructed inverse rigidity \(q/p\).
(b) Correlation between the \(q/p\) values reconstructed with the
VSP--GEM tracklets and CTS tracks; the slope of the linear fit is
consistent with unity.}
\label{fig:vsp_qp_performance}
\end{figure}

Figure~\ref{fig:vsp_qp_performance}(a) shows the reconstructed \(q/p\)
distribution. For validation, the VSP--GEM tracklets were matched to CTS tracks
reconstructed in the same events. Figure~\ref{fig:vsp_qp_performance}(b) shows
the correlation between the two \(q/p\) estimates. A linear fit to the
correlation gives a slope consistent with unity.

This demonstrates that VSP--GEM tracklets provide a reliable inverse-rigidity
estimate in the region where they overlap with the CTS. Their larger
geometrical acceptance makes this approach promising for extending track
reconstruction towards the phase-space region not fully covered by the CTS.
\section{Conclusion}
\label{sec:conclusions}

The Vertex Silicon Plane was commissioned and operated for the first time
during BM@N Run~9 as the tracking station closest to the target. Successful
integration of the free-streaming VSP data into the BM@N data stream was
demonstrated by the time coincidence of VSP hits with the trigger and their
spatial matching with reconstructed tracks. The detector showed stable in situ
performance, with the most probable cluster SNR values ranging from 15.5 to
33.2 and an amplitude-corrected hit-time distribution with a width of about
\(12~\mathrm{ns}\). In the nominal magnetic field of the SP-41 magnet, the
Lorentz drift of charge carriers leads to a displacement of the reconstructed
cluster position. For module~5, the measured charge-centroid shift is about
\(9~\mu\mathrm{m}\) on the n-side, while the corresponding p-side shift is
compatible with zero within the present uncertainty. The measured Lorentz
angle provides the basis for correcting this displacement in the reconstructed
hit position.

The tracking performance was studied using independent FSD- and GEM-based
reference tracks. Both reference samples give consistent VSP hit-matching
efficiencies of about \(95\%\). The track--hit residuals show good agreement
between VSP hits and the extrapolated reference tracks.

The contribution of the VSP to the performance of the CTS was also studied. It was shown that the VSP improves the transverse
track impact-parameter resolution, particularly in the \(x\) (bending)
projection. Monte Carlo studies show that the VSP--GEM tracklet topology
extends the geometrical reconstructability beyond the CTS acceptance. For
primary protons with \(p_T>1.2~\mathrm{GeV}/c\) in the 0--30\% centrality
class, the gain is about \(8.5\) percentage points, corresponding to an
increase of about \(24\%\) relative to the CTS. In the physics data, VSP--GEM
tracklets provide an inverse-rigidity estimate consistent with that obtained
from CTS tracks.

Thus, the VSP not only operated successfully as an additional silicon tracking
station, but also extends the BM@N tracking capabilities towards larger polar
angles and the midrapidity region. Further Monte Carlo reconstruction studies
are required to quantify the tracking efficiency in the extended acceptance
and to demonstrate the full reconstruction capabilities of the VSP--GEM
tracklet approach.

\section*{Author contributions}

D.~Dementev: conceptualization, project coordination, methodology, formal
analysis, data curation, detector commissioning and operation during Run~9,
supervision, writing -- original draft.

M.~Shitenkov: readout electronics, calibration, electronics testing, detector
commissioning and operation during Run~9.

A.~Kolozhvari: software, data curation, integration of the VSP data stream into
the BM@N event builder, detector commissioning and operation during Run~9.

V.~Leontyev: online data monitoring, data processing, detector commissioning
and operation during Run~9.

A.~Sheremetev: assembly of the silicon detector modules and the VSP station,
installation and detector commissioning.

D.~Andreev: methodology, mechanical design of the VSP station, installation
and detector commissioning.

R.~Arteche Diaz: firmware development and readout synchronization for the VSP.

D.~Baranov: software, development of the VSP hit producer and Monte Carlo
detector model.

C.~Ceballos Sanchez: writing -- review and editing.

I.~Gabdrakhmanov: software, integration of VSP monitoring histograms into
the BM@N online monitoring system.

I.~Gorelikov, S.~Reyes Pe\~na and E.~Tsapulina contributed to detector
operation during Run~9.

T.~Lygdenova: installation, detector commissioning and operation during Run~9.

Yu.~Murin: project administration, supervision.

I.~Osokin: slow-control system.

I.~Rufanov: methodology, formal analysis, alignment studies,
writing -- review and editing.

A.~Voronin: fabrication of carbon-fiber structural components, detector
commissioning and operation during Run~9.

\section*{Acknowledgements}

The authors thank Sergey Novozhilov and Semyon Piyadin, Technical Coordinator
of the BM@N experiment, for their assistance during the installation and
commissioning of the VSP.

The authors are grateful to the BM@N DAQ team for their support and for
providing equipment used during the development and debugging of the VSP
integration into the BM@N data-acquisition system.

The authors also thank Konstantin Gertsenberger, Alexandr Zinchenko,
Nikolay Zamyatin, and Mikhail Kapishin for valuable discussions, as well as
the BM@N software and data-analysis group for support with reconstruction
and analysis software.

The JINR accelerator team is acknowledged for their support during Run~9
operation.

\section*{Funding}
This work was supported by the Joint Institute for Nuclear Research (JINR) and the BM@N collaboration. 

\section*{Conflicts of Interest}

The authors declare that they have no conflicts of interest.

\section*{Additional Information}
\noindent\textbf{Use of artificial intelligence.}
During the preparation of this manuscript, the authors used OpenAI for language editing, improvement of text structure, and stylistic revision. All scientific content, experimental results, data analysis, and conclusions were produced and verified by the authors. The AI-assisted text was reviewed and edited by the authors, who take full responsibility for the content of the article.

\end{document}